\documentclass[lettersize,journal]{IEEEtran}
\usepackage{amsmath,amsfonts}
\usepackage{algorithmic}
\usepackage{array}
\usepackage[caption=false,font=normalsize,labelfont=sf,textfont=sf]{subfig}
\usepackage{textcomp}
\usepackage{stfloats}
\usepackage[labelsep=period,bf]{caption}
\usepackage{subcaption}
\usepackage{parskip}
\usepackage{placeins}
\usepackage[export]{adjustbox}

\usepackage{booktabs}
\usepackage{colortbl}
\usepackage{url}
\usepackage{verbatim}
\usepackage{graphicx}
\usepackage{xcolor}
\usepackage{hyperref}
\def\BibTeX{{\rm B\kern-.05em{\sc i\kern-.025em b}\kern-.08em
    T\kern-.1667em\lower.7ex\hbox{E}\kern-.125emX}}
\usepackage{balance}
\begin{document}
\title{Illuminating the Path: Attention-Assisted Beamforming and Predictive Insights in 5G NR Systems}

\author{Dino Pjanić, \IEEEmembership{Member, IEEE}, Guoda Tian, \IEEEmembership{Student, IEEE}, \\ Andres Reial, \IEEEmembership{Senior Member, IEEE}, Xuesong Cai, \IEEEmembership{Senior Member, IEEE}, \\Bo Bernhardsson,  and Fredrik Tufvesson, \IEEEmembership{Fellow, IEEE}\\
\thanks{Dino Pjanić, is with Ericsson AB, Lund, Sweden (e-mail: dino.pjanic@ericsson.com) and the Department of Electrical and Information Technology, Lund University, Lund, Sweden. (e-mail: dino.pjanic@eit.lth.se)  (\textit{Corresponding author: Dino Pjanic}) \par
Guoda Tian and Andres Reial are with Ericsson AB, Lund, Sweden. (e-mail: guoda.tian, andres.reial@ericsson.com) \par
X. Cai is with the School of Electronics, Peking University,
Beijing, 100871, China (email: xuesong.cai@pku.edu.cn) and the Department
of Electrical and Information Technology, Lund University, Lund, Sweden (email: xuesong.cai@eit.lth.se). \par
Bo Bernhardsson is with the Department of Automatic Control, Lund University, Lund, Sweden. (e-mail: bo.bernhardsson@control.lth.se) \par
Fredrik Tufvesson is with the Department of Electrical and Information Technology, Lund University, Lund, Sweden. (e-mail: fredrik.tufvesson@eit.lth.se) \par
Dino Pjanić and Guoda Tian contributed equally to this work.\par

The work is partially sponsored by the Swedish Foundation for Strategic Research and Ericsson AB, Sweden.}}


\markboth{}{Pjanić {et al.}: Illuminating the Path: Attention-Assisted Beamforming and Predictive Insights in 5G NR Systems}

\maketitle

\begin{abstract}
Artificial intelligence advances have recently influenced wireless communications, including beam management in fifth-generation (5G) new radio systems. AI-driven models and algorithms are being applied to enhance tasks such as beam selection, prediction, and refinement by leveraging real-time and historical data. These approaches address challenges such as mobility under complex channel conditions, showing promising results compared to traditional methods. Beam management in 5G refers to processes that ensure optimal alignment between the base station and user equipment for effective signal transmission and reception based on real-time channel state information and user positioning. This study leverages accurate beam prediction to identify a smaller subset of beams, resulting in a more efficient, streamlined, and link-adaptive communication system. The innovative approach presented introduces a precise, attention-based prediction model that derives the entire downlink transmission chain in a commercial grade 5G system. The predicted downlink beams are specifically tailored to handle the complexities of none line-of-sight environments known for high-dimensional channel dynamics and scatterer-induced signal variations. This novel method introduces a paradigm shift in utilizing environmental and channel dynamics in contrast to conventional procedures of beam management, which entails complex methods involving exhaustive techniques to predict the best beams. The presented beam prediction results demonstrate robustness in addressing the challenges posed by signal-dispersive environments, showcasing great potential in mobility scenarios.
\end{abstract}


\section{Introduction}
\label{Intro} 
\IEEEPARstart{B}{eamforming} is a signal processing method that directs radio energy through the channel toward a targeted receiver. Massive multiple-input multiple-output (MIMO) is an advanced antenna technology that provides high flexibility in beamforming due to the many radio frequency chains it employs \cite{LarsssonOveFredrik}. By adjusting the phases and amplitudes, the system can create constructive interference in the desired area and destructive interference in others. This approach enables focused beams toward specific receivers, enhancing signal strength and providing greater spatial diversity and multiple data streams. To sustain effective beamforming, especially with moving users, the system requires precise channel state information (CSI), which reflects the current characteristics of the communication channel between the transmitter and receiver.
Beamforming performs best in line-of-sight (LoS) scenarios, where there is an unobstructed path between the transmitter and receiver. In contrast, none LoS (NLoS) scenarios, such as urban environments with numerous obstructions, present significant challenges. In these cases, as users or obstacles move, signals often reflect off surfaces, resulting in multipath propagation, which demands advanced algorithms and precise CSI processing to dynamically match the instantaneous multipath propagation. This makes it particularly difficult to maintain accurate real-time CSI estimates for multiple beams, especially in scenarios involving high-mobility users such as vehicles. In NLoS environments with rapid changes or where CSI is noisy or incomplete, a reduced beam set allows the system to focus on the most reliable beams or those contributing the most signal energy, rather than trying to support numerous weak or scattered beams. Prediction of the strongest beams and beam reduction are closely interrelated, as both help to improve 5G beam management (BM) \cite{38802}. They are recognized as resource optimization strategies in MIMO systems, aiming to minimize the number of active beams or spatial streams employed during transmission or reception. \newline 
\indent Recently, advances in machine learning (ML) and artificial intelligence (AI), particularly deep neural networks (DNNs) such as transformer models introduced \cite{AttentionIsAllUneed}, have emerged as powerful tools to tackle a wide range of tasks. Originating in natural language processing (NLP), transformers utilize a unique mechanism known as self-attention, enabling them to capture long-term dependencies more effectively than traditional recurrent neural networks (RNNs). This makes transformers especially suitable for analyzing long sequences in time series data, such as CSI measurements influenced by UE mobility patterns and surrounding scattering characteristics in wireless environments. In this paper we study beam prediction in 5G NR systems based on attention models and channel fingerprints.\newline
\indent The rest of the article is organized as follows: the next section outlines the motivation for this study and provides an overview of key ML/AI applications in BM within 5G communication systems. Subsequently, we propose a transformer architecture with an attention-based model tailored for BM, with a focus on downlink beam predictions. Finally, we evaluate the performance of the proposed model with a particular emphasis on long-term beam prediction in demanding NLoS environments.

\section{Motivation}
\label{Motivation}
In wireless systems, coherence time refers to the period during which the channel impulse response or the transfer function, both with respect to phase and amplitude, remain relatively stable. Hence, CSI acquisition must be processed at millisecond level to track channel dynamics during mobility under varying environmental conditions. In legacy BM techniques, primarily employed in millimeter-wave (mmWave) communications, the base station (BS) transmits reference signals (RSs) and configures UEs to measure and report them. These measurements, along with associated reporting, impose significant overhead, a challenge that becomes particularly pronounced in dynamic and dispersive NLoS scenarios. Predictive methods, such as AI-assisted beam predictions, present a promising solution to reduce the reliance on continuous RS transmissions and measurements, addressing the limitations of traditional parametric models and solutions to meet the required capacity and performance improvements \cite{Gunduz}. Furthermore, conventional mathematical approaches to beam management often rely on idealized assumptions, such as pure additive white Gaussian noise, which may not accurately reflect real-world conditions \cite{Lajos}, creating opportunities for AI/ML models to capture and model complex nonlinear factors effectively. Advances in AI and ML have introduced a transformative perspective to 5G NR standardization \cite{38908} \cite{22874}, particularly through Release-18 \cite{38843}, which explores AI/ML-driven approaches to address scalability issues of MIMO systems such as increased antenna array sizes. The authors of \cite{StandardOverview} and \cite{StandardOverview2} provide an overview of current standards in relation to AI / related to AI / ML techniques. Many of the suggested methods replace traditional sequential beam sweep with predictive algorithms operating in the temporal and spatial domains; detailed insights are provided in \cite{AIML_BM} - \cite{AIML_BM4}. Key use cases, such as CSI feedback, beam management, and positioning, capitalize on the channel's temporal stability within the coherence time to reduce complexity. Recent research demonstrates the feasibility of short-term beam predictions using variations in the angles of arrival (AoA) and departure (AoD) in mobile environments. Hybrid approaches \cite{Alkhateeb} - \cite{LowFreqCSI_Access} leverage prior low-frequency channel information to predict optimal mmWave beams, reducing training overhead. In general, these approaches illustrate how the use of spatial channel characteristics in the sub-6 GHz band can simplify the complexity of the mmWave beam prediction encountered at mmWave frequencies. Different cross-domain approaches are also proposed in \cite{Lidar}, using LidarDAR sensors to improve time-domain beam prediction, while the authors of \cite{Radar} employ multipoint radar sensing to enhance beam tracking.\newline
\indent However, most recent BM studies rely on simplified LoS-dominant scenarios, often using simulated data as input of AI/ML models. In LoS environments, the best beam generally remains stable, suggesting that beam predictions could feasibly extend beyond coherence time if the UE follows a predictable movement trajectory. Another significant aspect is the fact that short-term prediction faces limitations: coherence regions typically span only a few decimeters or centimeters, depending on the frequency band, restricting prediction to brief intervals. We foresee advantages in longer-term beam predictions, where fingerprinting approaches can be explored by leveraging spatial and historical data patterns that indirectly incorporate AoA and AoD information through high-dimensional features. While fingerprinting is well-suited for stable and predictable radio channel environments, it requires robust augmentation with adaptive algorithms to handle the complexities of NLoS and dynamic scenarios effectively. However, in dispersive, NLoS environments even when the UE's trajectory is approximately known, the optimal beam prediction becomes highly sensitive to precise UE locations, often down to fractions of a wavelength.  
\indent 
Since the UE position inherently correlates with the best beam, this spatial information becomes a key factor in our prediction. The proposed attention-based solution adaptively learns the features of the measured UL channel and captures the features of the propagation characteristics, such as UE locations and surrounding environmental structures. In subsequent sections, we compare the energy efficiency based on the beam predictions of the proposed attention-based model. This involves evaluating the total energy of the full beam array against the energy represented by the predicted subset, providing insights into how efficiently the subset captures the beam energy relative to the entire array.

\subsection{Contributions}
\begin{itemize}
\item We present an accurate attention-based model for beam prediction that utilizes UL channel estimates from a commercial 5G system to derive the entire DL transmission chain, specifically tailored to handle the complexities of NLoS environments.
\item The novel approach introduced here enables accurate beam prediction far beyond the coherence time by utilizing high-dimensional fingerprinted features. These features predict temporal changes in the AoA and AoD, offering a more robust solution for dynamic and complex wireless environments.
\end{itemize}

\section{System Model and Problem Formulation}
\label{System_model}
In our single-user massive MIMO setup, the BS uses orthogonal frequency division multiplexing (OFDM) with $F$ subcarriers. At time $t$, the UE transmits an SRS pilot signal using $M_{UE}$ antennas. The BS has $M_{BS}$ antennas, evenly split between vertical and horizontal polarization.  Furthermore, let $P$ denote the number of multipath components, $\tau_{p,t}$ represent the time delay for the $p$-th path between the UE and BS, and $\alpha_{p,m,t}$ indicate the complex coefficient of the $p$-th multipath component at time $t$. UE transmits a pilot signal that reaches the BS antenna array at an azimuth arrival angle $\phi_p$ and an elevation angle $\theta_p$ for the $p$-th multipath component, respectively.
All vertically polarized antennas are used to form $M_{bm}^{Vt}$ beams, with the response of the $i$-th beam given by $\beta_{V,i}(\phi_p, \theta_p, f)$, where $f$ represents the pilot signal frequency. Similarly, horizontally polarized antennas generate $M_{bm}^{Hz}$ beams, with the response of the $i$-th beam defined as
$\beta_{H,i}(\phi_p, \theta_p,f)$. The total number of beams is $M_{bm} = M_{bm}^{Vt} +  M_{bm}^{Hz}$.
Consequently, for the $m$-th UE antenna the propagation channel model at time $t$ for each beam is:
\begin{equation}
\begin{aligned}
     h_{\mathbf{V},i,m,t}(f) &= \sum_{p=1}^P \beta_{V,i}(\phi_p, \theta_p, f)\hspace{1pt}\alpha_{p,m,t}\hspace{1pt} e^{-j(2\pi\hspace{1pt}f\hspace{1pt}\tau_{p,t})}  \\
     h_{\mathbf{H},i,m,t}(f) &= \sum_{p=1}^P \beta_{H,i}(\phi_p, \theta_p, f)\hspace{1pt}\alpha_{p,m,t}\hspace{1pt} e^{-j(2\pi\hspace{1pt}f\hspace{1pt}\tau_{p,t})}.
\end{aligned}
\end{equation}
By aggregating $h_{\mathbf{V},i,m,t}(f)$ and $h_{\mathbf{H},i,m,t}(f)$ from the $F$ subcarriers, we construct two channel transfer functions (CTF), 
$\mathbf{H}_{\mathbf{V},m,t} \in \mathbb{C}^{M_{bm}^{Vt} \times F}$ and $\mathbf{H}_{\mathbf{H},m,t} \in \mathbb{C}^{M_{bm}^{Hz} \times F}$ representing the vertical and horizontal polarized antennas, respectively, at time $t$. These matrices are strongly influenced by the UE's position, making them effective raw channel fingerprints for predicting DL beamforming weights, as the beam direction explicitly depends on the UE's location and implicitly on the CSI. Finally, for all UE antennas, the combined channel matrice is $\mathbf{H}_{UL,t} \in \mathbb{C}^{N \times F} = \left[\mathbf{H}_{\text{H},1,t}^T,\mathbf{H}_{\text{V},1,t}^T,..., \mathbf{H}_{\text{H},M_{\text{UE}},t}^T,\mathbf{H}_{\text{V},M_{\text{UE}},t}^T\right]^T$, where $N = M_{UE}\hspace{1pt}M_{bm}$. In the subsequent stage of the signal processing chain in a time division duplex (TDD) system, the beamforming weights are determined using the minimum mean square error (MMSE) channel estimator algorithm (\ref{MMSE}). These algorithms are designed to optimize the signal strength in the desired direction while minimizing interference. In a system with $M_{BS}$ antennas, the DL CTF can be estimated using the complex conjugate of the UL channel matrix, exploiting the reciprocity principle inherent in TDD systems, $\mathbf{\hat H}_{DL} = \mathbf{H}_{UL}^{*}$.
\begin{equation}
\begin{aligned}
     \mathbf{\hat H}_{\text{DL}} &= (\mathbf{H}_{\text{UL}}^{H}\mathbf{H}_{\text{UL}}+\sigma^2\mathbf{I})^{-1}\mathbf{H}_{\text{UL}}^{H}.
\end{aligned}\
\label{MMSE}
\end{equation}
where $\mathbf{H}_{UL}^{H}$ denotes conjugate transpose of the channel matrix, $\sigma^2$ noise variance, and $\mathbf{I}$ identity matrix.

The MMSE calculation detailed in \eqref{MMSE} is computationally intensive. In practice, only a small subset of beamforming weights in the matrix $\mathbf{H}_{DL}$ constitute beams that account for the majority of the beamforming energy. Predicting the strongest beams for future time instances can help reduce both energy consumption and computational load. However, it is challenging to accurately predict $\mathbf{H}_{UL}$ and eventually $\mathbf{H}_{DL}$ for future time instances, especially when the time gap exceeds the channel coherence time. This is because the further the prediction is, the more difficult it becomes to model the channel dynamics with high precision.
This drives the need to explore non-traditional approaches, such as AI-based algorithms, to predict the strongest beams in advance.
The proposed attention-based model learns a functional relationship $\boldsymbol{\psi} = f(\mathbf{H}_{UL,t}, \Delta_t)$, where $\boldsymbol{\psi} \in \mathbb{R}^n$ represents the $n$ strongest beams in the predicted matrix $\hat{\mathbf{H}}_{DL}$, and $\Delta_t$ denotes the time difference between current and future snapshots.


\section{Methodology}
\label{methodology}
CSI acquisition in TDD systems can benefit from channel reciprocity, meaning that the uplink (UL) and downlink (DL) channels are related since both use the same frequency band. This allows for estimating the CSI on the UL and applying it to the DL beamforming. For example, a BS equipped with 64 transceivers can leverage a single uplink pilot to estimate the full 64-dimensional channel across the entire bandwidth. This is facilitated by the transmission of sounding reference signals (SRS) from the multiantenna UE, enabling the BS to estimate the DL channel and compute DL beamforming weights (BFWs). These weights, represented as complex coefficients, are applied to the BS's MIMO antenna elements to control signal direction and shape by adjusting the amplitude and phase of the DL signal.

\subsection{Dataset collection}
\label{datacollection}
\begin{figure}[b!]
    \centering
    \centerline{\includegraphics[width=\columnwidth]{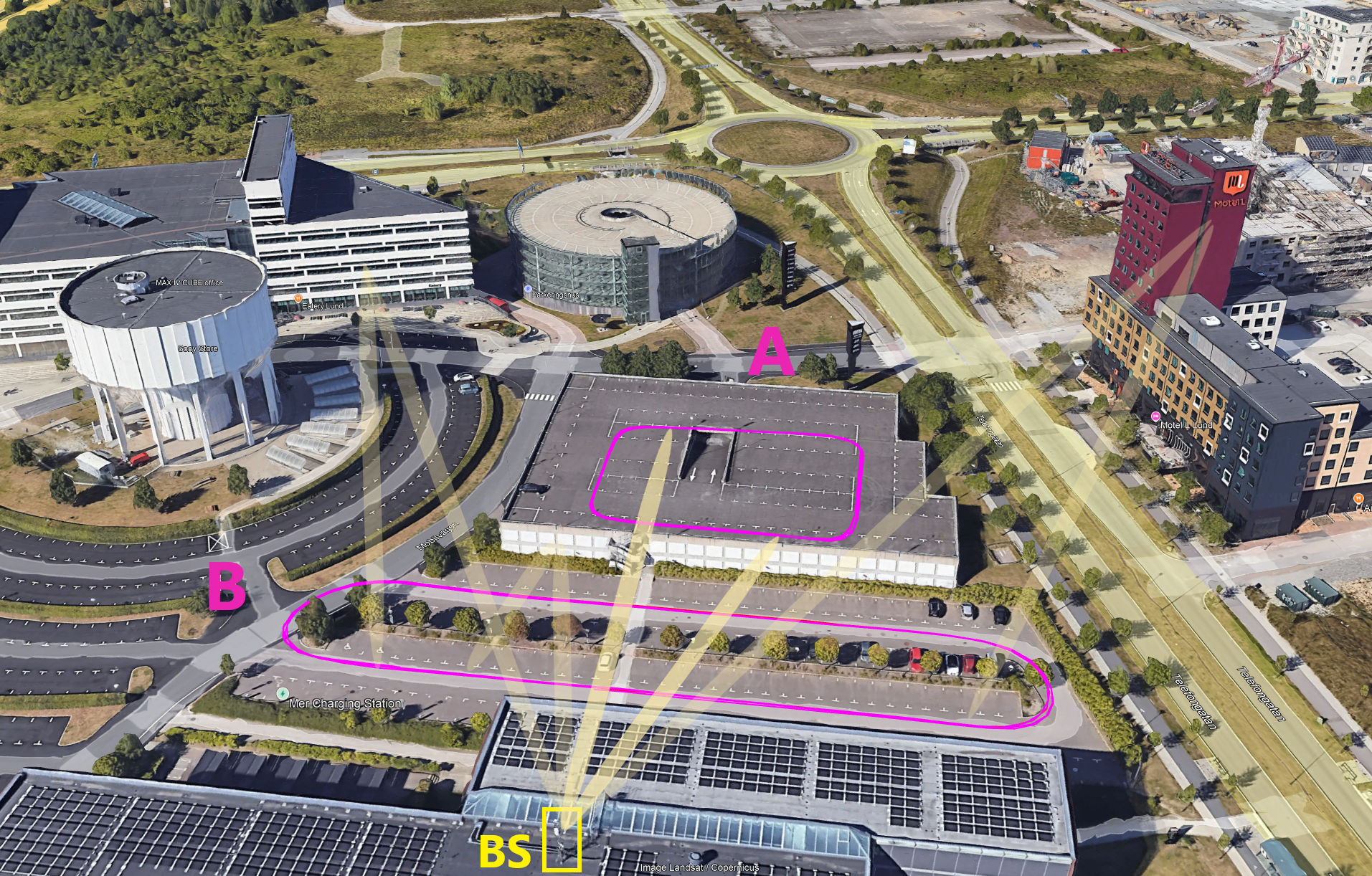}}
    \caption{A BS placed at a 20 m high rooftop. SU-MIMO scenario was tested in the two pre-defined measurement routes: \textbf{A}: The roof of a 10 m high garage building was used for LoS propagation measurements with a strong dominant path along the entire route. \textbf{B}: A ground-level route where the signals reflect off multiple surfaces and surrounding buildings block the signal causing NLoS propagation.}
    \label{fig:test_area}
\end{figure}
To enable TDD-based downlink beam prediction using UL SRS channel estimates, a commercial grade 5G BS was used, compliant with 3GPP standards \cite{Toskala} - \cite{38901} for radio resource management, physical layer considerations and beam measurement procedures. The BS operated at a center frequency of 3.85 GHz with a bandwidth of 100 MHz and was equipped with a rooftop-mounted phased array antenna module (PAAM) comprising 64 cross-polarized antenna elements. A 5G-capable UE was kept connected while simultaneously downloading data at a sustained rate of 750 Mbit/s to maintain continuous UL SRS transmission throughout the measurements. The baseband unit of the BS processed the time-varying SRS reports to extract channel estimates, operating initially in the antenna element domain. The SRS data received, represented as complex samples, underwent additional unpacking from 16-bit floating-point format to 2xSQ15. The pre-processing steps included undoing the normalization of the SRS channel estimates averaged over the physical resource blocks (PRB) pairs to derive estimates for each PRB and beam. Finally, the processed SRS samples were transformed into the beamspace domain using a fast Fourier transform (FFT), producing a time series of beam measurements. \newline 
\indent The measurement campaign encompassed two distinct propagation scenarios, LoS and NLoS, as illustrated in Fig. \ref{fig:test_area}, as these provided a range of challenging environments to evaluate the proposed approach. Data collection was performed on these two approximately rectangular routes. Each route was repeated over five laps, serving as baselines for analysis. All measurements were performed using a test vehicle moving at a steady velocity of 15 km/h (4.2 m/s). The approach of comparing two fundamentally opposite propagation scenarios serves to establish a simplified baseline with the LoS scenario while contrasting it with the more complex NLoS scenario. The NLoS scenario introduces a non-trivial relationship between the UE trajectory and the scatterers, collectively influencing the optimal beam direction. The prediction method relies on repetitive UE movement paths, emulating a car traveling along a road or a robot/machine following a specific route in a factory setting. These scenarios provide consistent movement patterns that can be effectively captured during model training and utilized during inference, allowing the model to represent the dominant candidate beam paths. However, it is important to note that the optimal beam may still vary between different movement realizations during inference. However, in practice, most commercial site deployments exhibit predictable UE movement patterns due to the static nature of the surrounding environment, which further supports the feasibility of this approach.

\subsection{Signal Processing Framework}
\label{signalprocessing}
\begin{figure}[t!]
	\centering 
    \includegraphics[width=1\linewidth]{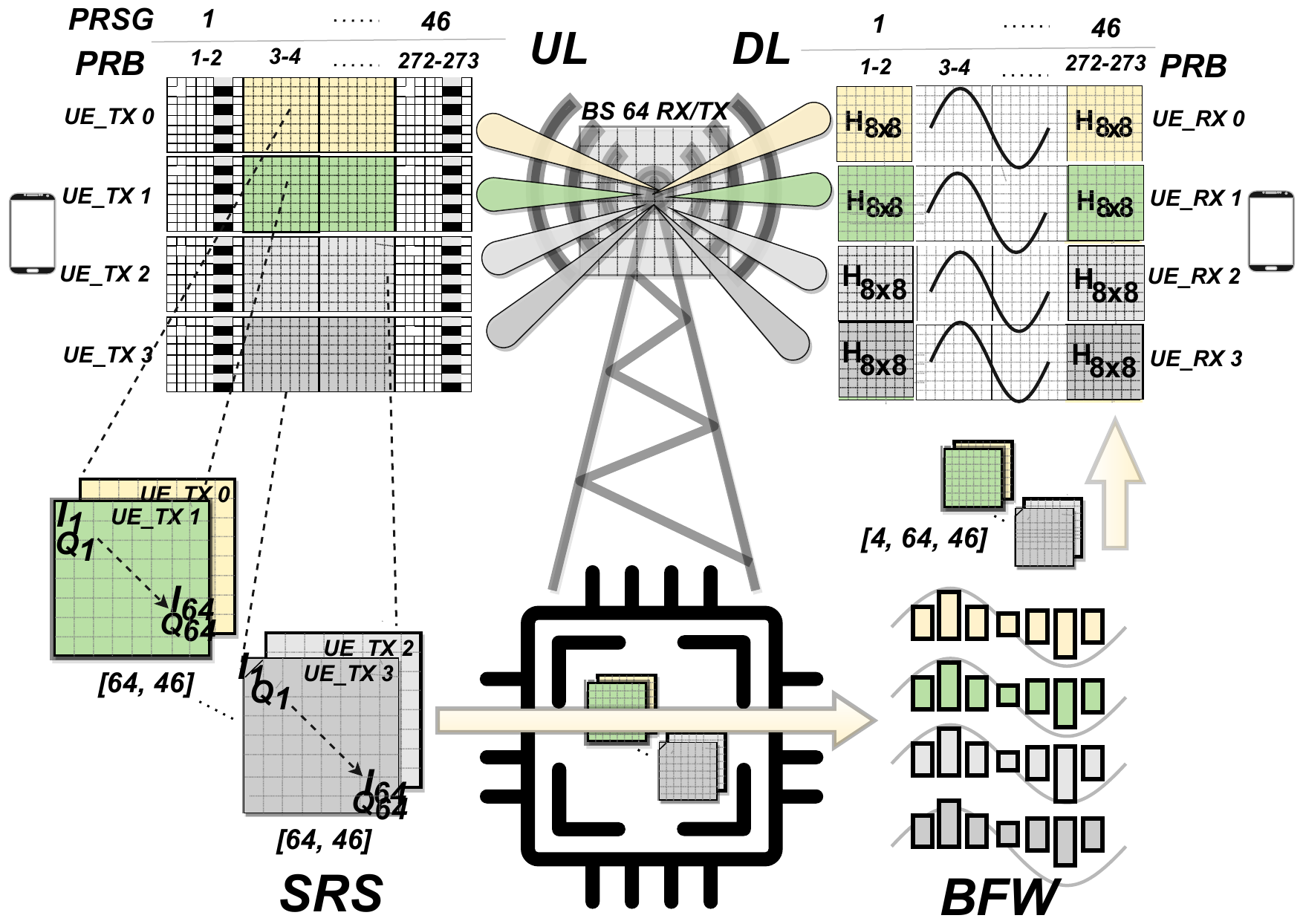}
	\caption{Beamforming weights generation based on SRS channel estimates.}
	\label{fig:srs-data-struct_with_BFW} 
\end{figure}

This section outlines the BS processing of the UL SRS channel estimates, which form the primary training data set. The BS handles a time series of SRS measurements, representing the angular delay spectrum of the radio channel in the beam domain. Simultaneously, a parallel computation determines the corresponding DL BFWs using MMSE-based algorithms, as described in Section \ref{System_model}. In this study, the UL SRS data serve as the input for the Attention-aided model, while the prediction task focuses on the generated DL beams, as detailed in the following chapters. As illustrated in Fig. \ref{fig:srs-data-struct_with_BFW}, the UL SRS channel estimates span 273 PRBs within a 100 MHz bandwidth. Each channel snapshot includes data for all 64 beams across these PRBs, based on an SRS reporting interval of 20 ms. To reduce complexity, the PRBs are grouped into adjacent pairs, with the average value of each pair calculated by downsampling. This process results in 137 PRB subgroups (PRSGs). Further refinement is achieved by grouping every three consecutive PRSGs, where the first PRSG is downsampled, and the second and third are interleaved. This approach ultimately yields 46 PRSGs. The UE, equipped with four antennas (corresponding to four UE layers), is responsible for transmitting the SRS pilot signals. The SRS pilots recorded from all four UE layers form CTF matrices $\mathbf{H}_1, \mathbf{H}_2, \mathbf{H}_3, \mathbf{H}_4 \in \mathbb{C}^{N \times F}$. The matrix $\mathbf{H}' \in \mathbb{C}^{4N \times F}$ contains all four matrices, specifically $\mathbf{H}' = [\mathbf{H}_1^T, \mathbf{H}_2^T, \mathbf{H}_3^T, \mathbf{H}_4^T]^T$. 
After further processing, the final 20-ms-based CTF snapshot is structured with 4 x [4 x 46] dimensions, representing amplitude instances. This data is collected for the 64 BS antennas and four UE layers in 46 PRSGs, as illustrated in Fig. \ref{fig:srs-data-struct_with_BFW}. \newline
\indent Obtaining uplink UL SRS channel measurements in a commercial 5G BS introduces significant challenges, particularly when working with large, complex data structures like SRS measurement samples. These measurements, generated at millisecond intervals, are typically confined to the baseband unit of the BS for internal operations, with external access often limited by hardware and software restrictions. Moreover, since not all PRSG values are updated during UL SRS transmissions, it becomes essential to account for and address missing channel estimate values.
To ensure the reliability of the collected UL CTF, the processing pipeline must include mechanisms to verify the validity of input data and handle incomplete PRSGs effectively. SRS channel estimates are classified as valid only if all PRSGs and UE transmit antennas have successfully transmitted SRS during a given discontinuous reception (DRX) cycle. Each PRSG comprises two PRBs, so channel estimates are derived by averaging the PRB pairs for every PRSG and beam.
Validation of the raw UL CTF matrix involves a two-step process. The matrix is deemed invalid if it meets any of the following criteria based on the number of missing subcarrier samples (represented as zero elements):
\begin{itemize}
    \item Insufficient CSI snapshots in the beam and frequency domain: the number of non-zero elements in $\mathbf{H}_t$ is lower than a given threshold of 60\%. 
    \item Updating procedure stalling: The values at all sub-carriers or all beams remain the same compared to the previous reporting interval.\\    
\end{itemize}
$\mathbf{Note}$: In cases where the UL CSI was determined to be insufficient, the calculation of DL BFWs was halted. \newline 
After discarding all invalid data, the next step is to process the raw UL CTF to generate impulse response beam matrices. To suppress the side lobes, we apply Hann windowing in all rows of the matrix $\mathbf{H}_t$ to obtain matrix $\hat{\mathbf{H}}_t \in \mathbb{C}^{N \times F}$. The  $F$-length Hann window in the frequency domain is given by:
\begin{equation}
    w[f] = \sin^2\left(\frac{\pi f}{F}\right), \quad f= 0, \ldots, F-1.
\end{equation}

After the windowing operation, the impulse response beam matrix $\mathbf{G}_t$ is produced by performing the inverse discrete Fourier transform along each row of $\hat{\mathbf{H}}_t$. Given the potential difficulty in achieving a stable phase for $\mathbf{G}_t $, here we opt to use its amplitude $|\mathbf{G}_t|$ as the training feature, although this means discarding potentially useful information. \newline 
\indent Digital beamforming involves the adjustment of antenna weights during the digital baseband processing. The independently calculated DL BFWs are applied to the downlink signal after transforming them from the beam domain to the antenna domain. In this process, each antenna in the array is assigned a specific phase and amplitude as determined by the computed weights.
\begin{figure}[htbp!]
	\centering  
    \includegraphics[width=0.93\linewidth]{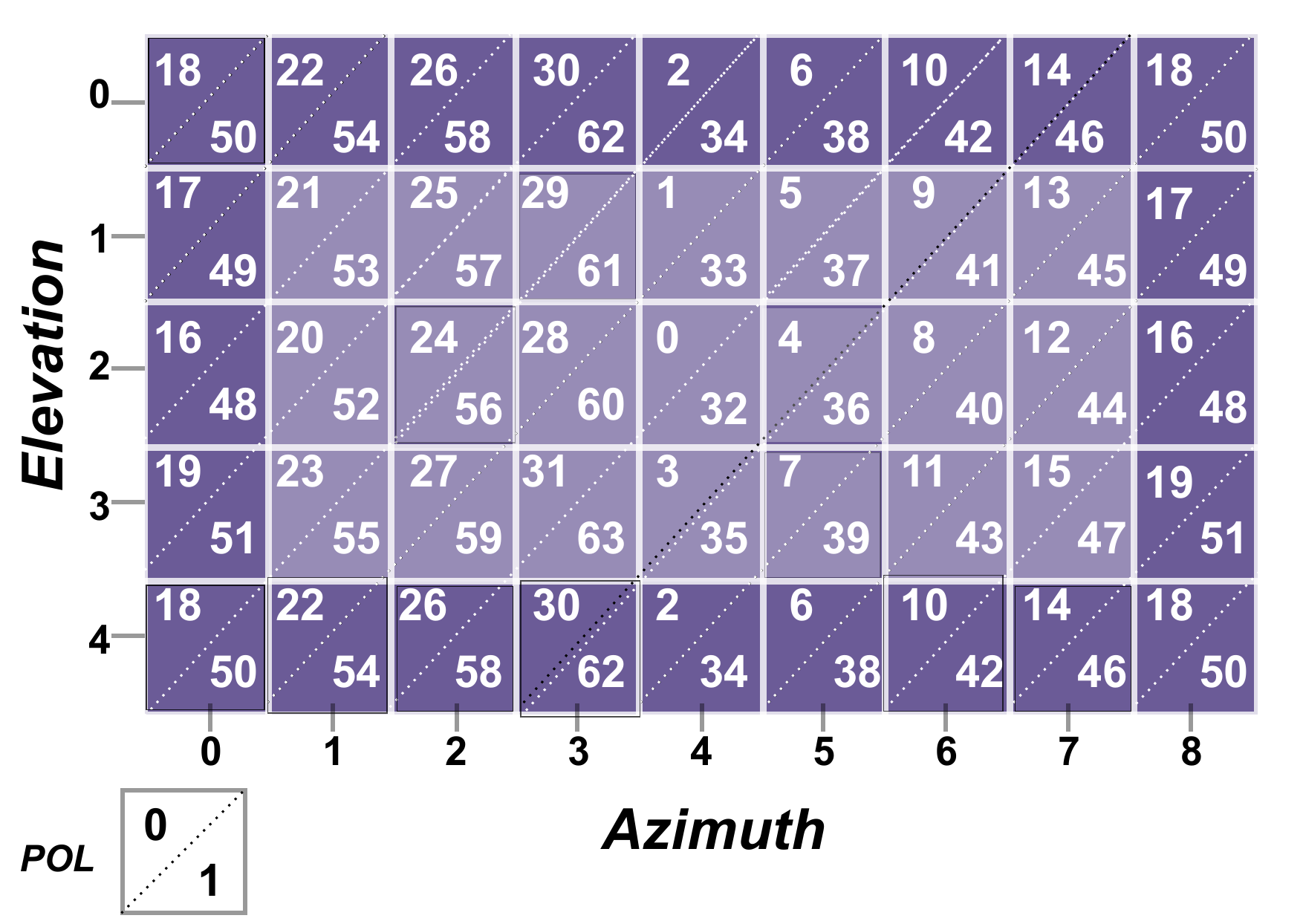}
	\caption{Grid of beams explaining the layout of beams produced by the deployed MIMO system, where each cell in the grid corresponds to a specific beam and its cross-polarized counterpart.}
	\label{fig:GOB} 
\end{figure}
The PAAM explained above generates a grid-of-beams (GoB), forming a structured set of beams that span the coverage area. Each beam represents a spatially focused transmission or reception pattern, enabling efficient signal delivery to or from specific UE locations.  As illustrated in Fig.~\ref{fig:GOB}, the GoB structure yields 64 distinct beams, providing precise control over the signal direction and maximizing spatial selectivity.

\section{Transformer Architecture} 
\label{Attention}
The transformer architecture, introduced in \cite{AttentionIsAllUneed}, has served as the foundation for numerous state-of-the-art models in natural language processing, thanks to its ability to effectively process input sequences and generate accurate output sequences. Unlike traditional methods that analyze tokens sequentially, transformers relate each token to all others within a sequence. Their self-attention mechanism replaces sequential processing with parallel computation, distinguishing them from recurrent neural networks (RNNs) \cite{RNN} and convolutional neural networks (CNNs) \cite{CNN}. Using parallelism, transformers efficiently capture long-range contexts and dependencies across distant positions in input or output sequences. \newline
\begin{figure}[t!]
	\centering 
    \includegraphics[width=1\linewidth]{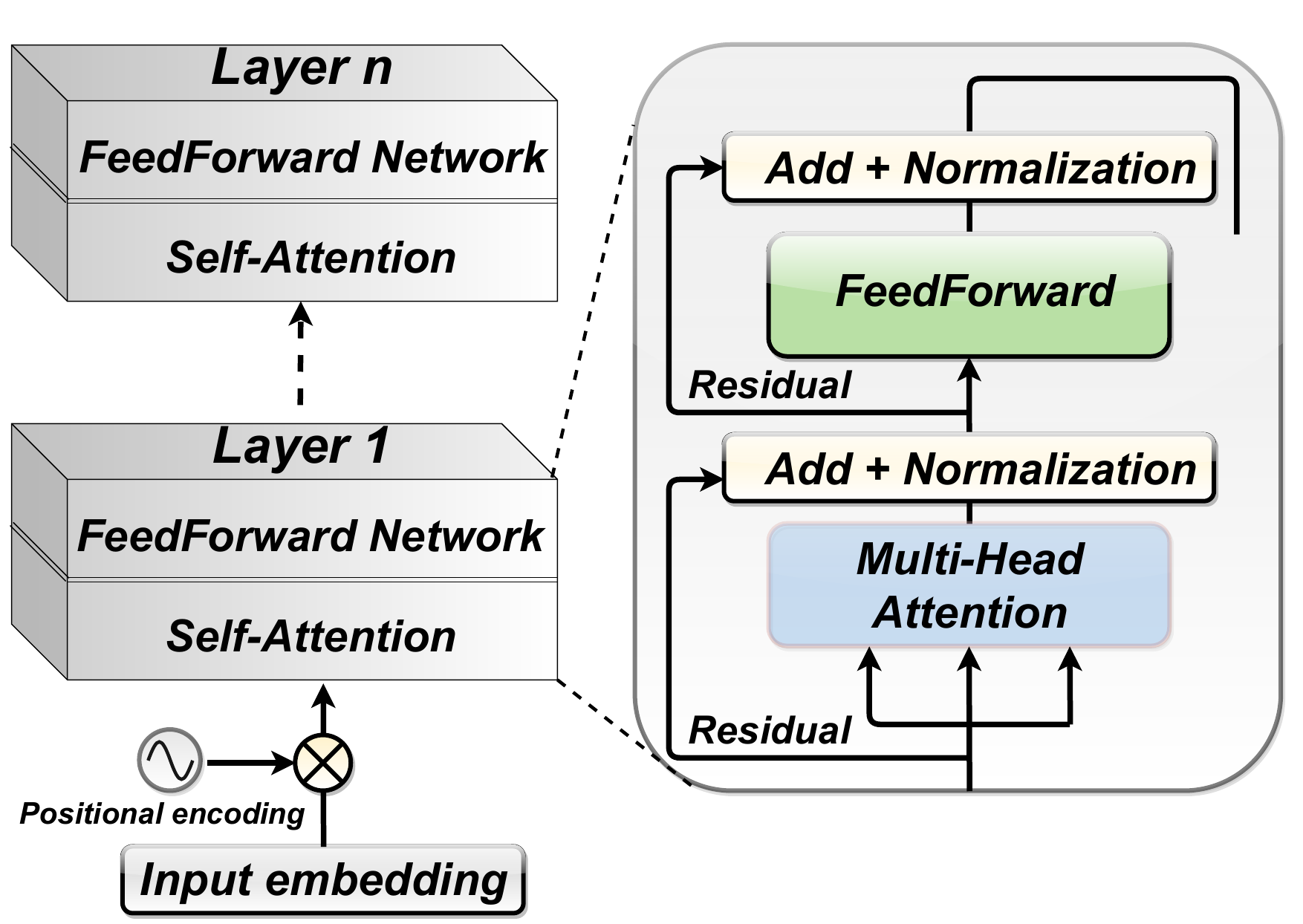}
	\caption{A stacked architecture of computational encoder layers. The proposed model deploys 3-layer architecture.}
	\label{fig:TransformerLayerSublayer} 
\end{figure}
\indent  The transformer is a Deep Neural Network (DNN) model \cite{DNN} composed of multiple layers with a uniform architecture. These layers are organized into stacks that differ from those in classical DNN models. Each stack, which can function as an encoder or a decoder, operates from bottom to top. The input and output sequences are transformed into vectors of dimension \emph{d} through embedding and positional encoding layers. Each layer sequentially passes its learned representations to the next layer until the final prediction is achieved. In particular, each layer comprises sublayers, all of which share an identical structure across different layers, enhancing hardware optimization. In its original design, the transformer includes two key sublayers: a self-attention sublayer and a feedforward network, as depicted in Fig. \ref{fig:TransformerLayerSublayer}. The self-attention sublayer is further divided into \emph{n} independent and identical components, called heads. The transformer architecture was originally designed for sequence-to-sequence tasks such as machine translation, and both encoder and decoder blocks were soon adapted as standalone models. Although there are hundreds of different transformer models, most of them belong to one of three types; encoder-only, decoder-only or encoder-decoder. In this study, we chose the encoder-only architecture to predict the best beams.

\subsection{Input Embedding}
The input embedding sublayer converts the input tokens to vectors of dimension $d\textsubscript{model}.$ Many embedding methods can be applied to the tokenized input. The later proposed model applies a simple lookup table that stores embeddings of a fixed dictionary and size. This module often stores word embeddings and retrieves them using indices. The input of the module is a list of indices, and the output is the corresponding word embeddings.

\subsection{Positional Encoding}
The idea behind positional encoding (PE) is to preserve sequential information in the input data sequence. This is achieved by adding value to the input embedding instead of having additional vectors to describe the position of a token in a sequence. Positional embedding provides sine and cosine functions that generate different frequencies for the PE for each entry $i$ of the $d\textsubscript{model}$ entries in the PE vector:
\begin{equation}
\begin{aligned}
    \mathbf{PE}(pos\:2i) &= \sin\left(\frac{pos}{{10000^{2i/d\textsubscript{model}}}}\right); \\
    \mathbf{PE}(pos\:2i+1) &= \cos\left(\frac{pos}{{10000^{2i/d\textsubscript{model}}}}\right).
\end{aligned}
\label{PosE}
\end{equation}
The sine function is applied to the even numbers and the cosine function to the odd numbers. These vectors follow a specific pattern that the model learns, which helps it determine the position of each token or the distance between different tokens in the sequence. In addition, positioning ensures meaningful distances between the embedding vectors once they are projected via dot-product operations in the attention mechanism.
\subsection{Encoder Stack}
As shown in Fig. \ref{fig:TransformerLayerSublayer}, the encoder consists of a stack of $n$ layers, each comprising two primary sublayers: a multihead self-attention block and a position-wise fully connected feed-forward network. To facilitate deeper models, a residual connection is applied around each of these sublayers, followed by layer normalization. Specifically, each sublayer, denoted as \texttt{sublayer(x)}, includes a residual connection that transports the raw input x of the sublayer directly to the normalization function of the layer. This ensures that critical information, such as positional encoding, is preserved throughout processing.
The output dimensions of all sublayers, as well as embeddings, are of size $d\textsubscript{model}$ which has a significant consequence, for example, all key operations are dot products. As a result, the dimensions remain stable, which reduces the number of operations.
\subsection{Self-Attention}
At the core of Transformers lies the self-attention mechanism. The input of a Transformer consists of a sequence of contiguous tokens, each represented as a vector in an embedding matrix. As part of the self-attention process, three projection matrices W\textsubscript{q}, W\textsubscript{k}, and W\textsubscript{v} transform each input embedding vector into three distinct vectors: the Query, Key, and Value.
For each token, its corresponding Key vector is compared to the Query vectors of all other tokens by computing dot products. This calculation provides a measure of similarity between Queries and Keys, forming the foundation of the attention mechanism. A Softmax function is then applied to normalize these similarity scores, amplifying the most relevant relationships. The softmax function is defined as below \eqref{Softmax}
\begin{equation}
\begin{aligned}
    {Attention}\mathbf{(Q, K, V)} = \text{softmax}\mathbf{\left(\frac{QK^\top}{\sqrt{d_k}}\right)}\mathbf{V}
\end{aligned}
\label{Softmax}
\end{equation}
\begin{figure*}[t!]
	\centering 
    \includegraphics[width=0.85\linewidth]{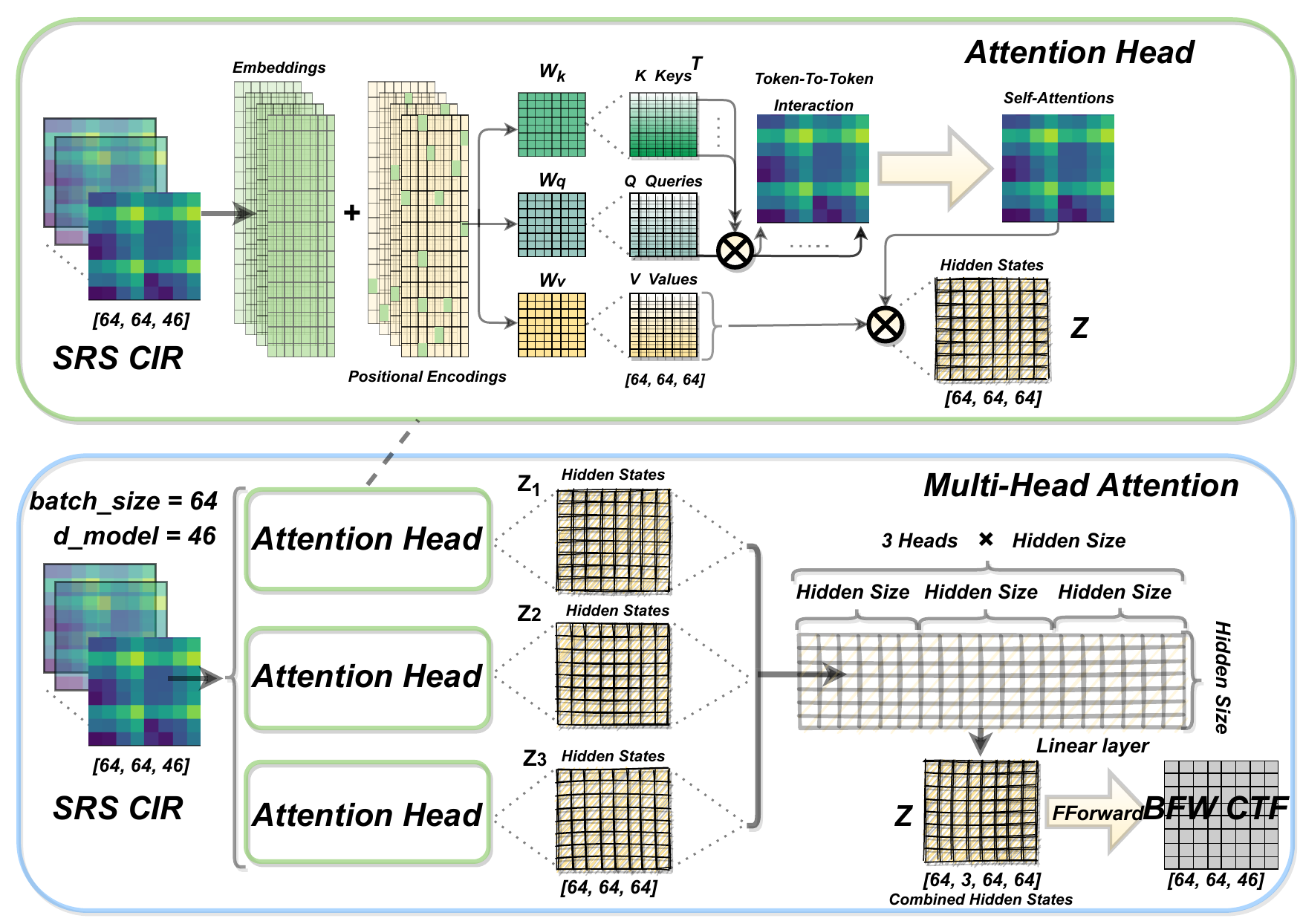}
	\caption{Attention model implementation and tensor dimensions.}
	\label{fig:MultiHeadAttention} 
\end{figure*}
where \textbf{Q}, \textbf{K} and \textbf{V} are representation matrices. In addition, $\boldsymbol{{d\textsubscript{k}}}$ represents the dimensionality of the Key vectors and $\sqrt{\boldsymbol{{d_k}}}$ is a scaling factor to prevent large values in the dot product. The softmax expression on the right-hand side of equation \eqref{Softmax} normalizes the similarity scores into probabilities.
The resulting matrix, known as the self-attention matrix, captures the contextual relationships between the tokens. This process is performed multiple times in parallel, using several independent sets of projections, resulting in a multi-head attention layer that enhances the model's ability to capture complex patterns across the input sequence.

\section{Proposed Transformer Framework} 
\label{MLModels}
The implemented Transformer model deploys a 3-layer, encoder-only architecture to predict the time series of DL beams based on UL SRS channel estimates. The paired data sets consist of the input UL channel impulse response (CIR) data, fed into the encoder, and the corresponding DL beam TF serving as the target dataset, described in detail in Section \ref{Attention}. The model is trained by minimizing the error between the encoder's output and the target DL beams. The input data, represented as the UL CIR beam matrix $\mathbf{G}_t \in \mathbb{C}^{N \times F}$, provides amplitude values used as input of a DNN equipped with an attention mechanism, while $\mathbf{B}_t \in \mathbb{C}^{N \times F}$ represents the DL TF. The attention-based model processes the input sequence to produce a rich numerical representation optimized for translating UL SRS channel estimation data into DL beam predictions. The architecture leverages bidirectional attention, where the representation of a given token considers both preceding and succeeding tokens in the sequence. This property is particularly well-suited for time-series data, as it captures the temporal dependencies in channel measurement sequences effectively.
The attention-based beam prediction pipeline, as described in Fig.~\ref{fig:MultiHeadAttention}, consists of an encoder-only deployment that comprises multiple attention-aided blocks, followed by an output layer that employs a loss functions, namely the Mean Square Error (MSE). We use $\boldsymbol{\eta}_i = [\eta_{bm_1,i}, ..., \eta_{bm_{64},i}]^T$ to represent the CTF of the DL beam as the ground truth of the moving UE at position $i$. This approach directly estimates the 64 DL beams by setting a regression head in the output layer of the last attention block. Let $f_{\text{MSE}}(.)$ denote the overall function and vector $\boldsymbol{\theta}_{2}$ all hyperparameters, $\boldsymbol{\eta}_i = [\eta_{bm_1,i}, ..., \eta_{bm_{64},i}]^T$ the estimated $i$-th 64-sized beam set generated by $f_{\text{MSE}}(\boldsymbol{\theta}_{2}, |\mathbf{G}_t|)$, the loss  $\boldsymbol\ell$ can be expressed as
\begin{equation}
    \boldsymbol\ell = \frac{1}{N_{train}}\sum_{i \in \Omega_{train}'}||\boldsymbol{\eta}_i-\hat {\boldsymbol{\eta}}_i||^2_F,
    \label{MSE}
\end{equation}
where $\Omega_{train}'$ and $N_{train}$ denote the training set and the number of training samples, respectively, and $||.||_F$ denotes the Frobenius matrix norm. \newline
\indent As illustrated in Fig.~\ref{fig:MultiHeadAttention}, an attention head operates in two key steps. First, as detailed in Section \ref{Attention}, the attention mechanism computes the keys \textbf{K} and queries \textbf{Q} from the input data. This process evaluates the relevance of each query vector \textbf{Q} with respect to all key vectors \textbf{K}, generating an energy score that reflects their importance. Or, simply explained how much attention should a token pay to another token in the input sequence. A Softmax transform normalizes and further accentuates the high similarities, and the resulting matrix is called self-attention. \newline
Next, the mechanism introduces a separate feature representation vector called values \textbf{V} that is combined with the attention weights (calculated from the dot products of \textbf{Q} and keys \textbf{K}) to produce the so-called hidden states. These hidden states represent a weighted sum of the values, highlighting the most relevant information. Notably, \textbf{Q}, \textbf{K} and \textbf{V} are all derived from the same input sequence, ensuring that the attention mechanism captures a rich and comprehensive representation of the data. The process is repeated multiple times with multiple attention layers, resulting in a multi-head attention layer. The final hidden states are combined into final hidden states by using a linear layer.
\vspace{-2pt}
\section{Results and Discussion}
\label{Results}
This section evaluates the AI-based prediction pipeline. We start by assessing the data cleaning process to ensure the retention of sufficient channel information. Next, we assess the beam energy by quantifying the predicted DL beamforming performance. This involves analyzing the energy distribution within the selected beam subsets across different prediction time horizons, providing information on the effectiveness of the prediction model in maintaining energy efficiency and accuracy over time.
\subsection{Single snapshot channel representation}
\begin{table}[b]
\centering
\caption{Architectural overview of the proposed encoder-only model.}
\vspace{-5pt}
{\begin{center}
\begin{tabular}{ccccccc}
\hline\hline
Model entity & Network Structures or Parameters  \\
\hline
Input Features & Amplitude of CIRs for all beams\\
Network Output & Estimated DL TF for all beams\\ 
Intermediate block 1 & Residual $3$-head Self-Attention Layer \\ 
Intermediate block 2 & Residual Position-wise FCNNs \\
Intermediate block 3 & $3$ cascaded ordinary FCNNs \\
Encoder layers & $3$ \\
d\textsubscript{model} & $46$  \\
Batch size & $64$\\
Optimizer & Adam \\
Learning Epochs  & $2500$\\
Time Complexity & $NF^2$\\
\hline\hline
\end{tabular}
\end{center}
}
\label{table: model_description}
\vspace{-10pt}
\end{table}
As detailed in Section \ref{datacollection}, the BS recorded channel snapshots for two propagation scenarios, LoS and NLoS, resulting in $\mathcal{T}{1} = 22000$ and $\mathcal{T}{2} = 24603$ snapshots, representing time instances, collected on a 20~ms time resolution basis. These snapshots are structured into two tensors, $\mathcal{A}{\text{LoS}} \in \mathbb{C}^{\mathcal{T}{1} \times N \times F}$ and $\mathcal{A}{\text{NLoS}} \in \mathbb{C}^{\mathcal{T}{2} \times N \times F}$, where each subset of four adjacent snapshots corresponds to the signals transmitted by four UE antennas (layers). Each tensor is normalized by multiplying with a scalar such that its Euclidean norm is equal to $\mathcal{T}_i\hspace{1pt}MN$, where $i = 1,2$. Following the validation of the input channel snapshot described in Section \ref{signalprocessing}, a single UL CTF matrix instance $\boldsymbol{\Xi}$ instance $\boldsymbol{\nu}$ applies a cut threshold of approximately $60\%$, yielding $1766$ of the $2944$ ($64 \times 46$) components, corresponding to 64 BS antennas in 46 PRSG. This ensures that adequate channel information is retained. The amplitude of the UL CIR beam matrix, $|\mathbf{G}_t|$, is then derived and passed to the attention-aided prediction block. The prediction block architecture, described in Table \ref{table: model_description}, comprises three cascaded sub-blocks, providing the foundation for downstream processing. Initially, position encoding is applied to $|\mathbf{G}_t|$ as described in (\ref{PosE}), followed by layer normalization. The normalized matrix is then fed into three parallel self-attention blocks, each comprising a single self-attention layer, as illustrated in Fig.~\ref{fig:MultiHeadAttention}, to produce the output matrix $\mathbf{Z}$ via (\ref{Softmax}). These multi-head attention layers process a sequence of size $\mathcal{T}$ (single snapshot), with each head projecting the feature dimensions $1766$ into smaller subspaces to compute the query \textbf{Q}, key \textbf{K}, and value \textbf{V} representations.
After the Add \& Normalization process, as depicted in Fig. \ref{fig:TransformerLayerSublayer}, the output is transferred to the second sub-block, consisting of two position-wise fully connected neural networks (FCNNs) with sizes $\mathbf{W}_1 \in 46 \times 64$ and $\mathbf{W}_2 \in 64 \times 46$. Following this, the output matrix of the second sub-block is vectorized to produce a vector of original length $2944$. This vector is fed into the last FCNN sub-block, with sizes defined in Table~\ref{table: model_description_detailed}. This entire computational process is repeated across all three layers of the encoder block, adhering to the architecture depicted in Fig. \ref{fig:TransformerLayerSublayer}. \newline
\indent As detailed in Section \ref{MLModels}, the model uses a paired dataset, where the input UL CIR data is processed by the encoder and the corresponding DL beam TF, $|\mathbf{B}_t|$, serves as the ground truth. 
The model training optimizes, via Adam optimizer, the encoder's output by minimizing the error relative to the target DL beam TF. The model's final predicted output represents the beam energy levels across all 46 PRSG subcarriers and 64 antennas of the BS. It encapsulates the total energy distribution across the frequency and spatial domains and reflects the combined radiated energy output of the beamforming system, integrating contributions from each antenna and subcarrier.

\begin{table}[t]
\centering
\caption{Overview of the last FCNN sub-block.}
\vspace{-5pt}
{\begin{center}
\begin{tabular}{ccccccc}
\hline\hline
Model entity & Dimensions  \\
\hline
Input layer size & $2944 \times 1$\\
Hidden layer 1 & $2944 \times 64$ \\ 
Hidden layer 2 & $64 \times 32$ \\
Hidden layer 3 & $64 \times 64$ \\
Cost function & \eqref{MSE}\\
\hline\hline
\end{tabular}
\end{center}
}
\label{table: model_description_detailed}
\vspace{-10pt}
\end{table}
\subsection{Beam Energy Evaluation Methodology}
The cumulative power across the predicted beam subset is the main subject of investigation to predict the total power expected in a series of predicted beams. In this study, instead of ranking the beams by their indices, we focus on selecting the strongest beams within each subset, meaning the indices of the predicted strongest beams may not always align with those in the ground truth dataset. Ranked by their expected power (e.g., strongest to weakest), the cumulative power is the sum of the power contributions of individual beams in a beam subset is defined as
\begin{equation}
\begin{aligned}
   \label{eqn:mape} 
     P_{\text{cumulative}}(n) = \sum_{i=1}^{n} P_i,
\end{aligned}
\end{equation}
where $P_{cumulative}(n)$ is the cumulative power of the first $n$ predicted beams, and $P_{i}$ is the power of the $i$-th beam. This cumulative metric can help quantify how well the predictions capture the total power available in the full set of beams, providing insights into how quickly the cumulative power saturates with increasing beams and helping determine the optimal subset size for energy-efficient operation.
When comparing the energy of the complete set with a subset, $\mathcal{N}$, to a subset $n$, where $n \in \mathcal{N}$, discrete subsets of beams are selected with sizes $n$ = [4,8,16,32]. Here, $\mathcal{N}$ represents the total energy from all beams within the antenna system, serving as a benchmark for evaluating the energy output. This approach enables a systematic assessment of the trade-offs between beam subset size and energy efficiency while maintaining the predictive accuracy of the system.
\begin{figure}[htbp!]
    \centering
    \centerline{\includegraphics[width=0.9\columnwidth]{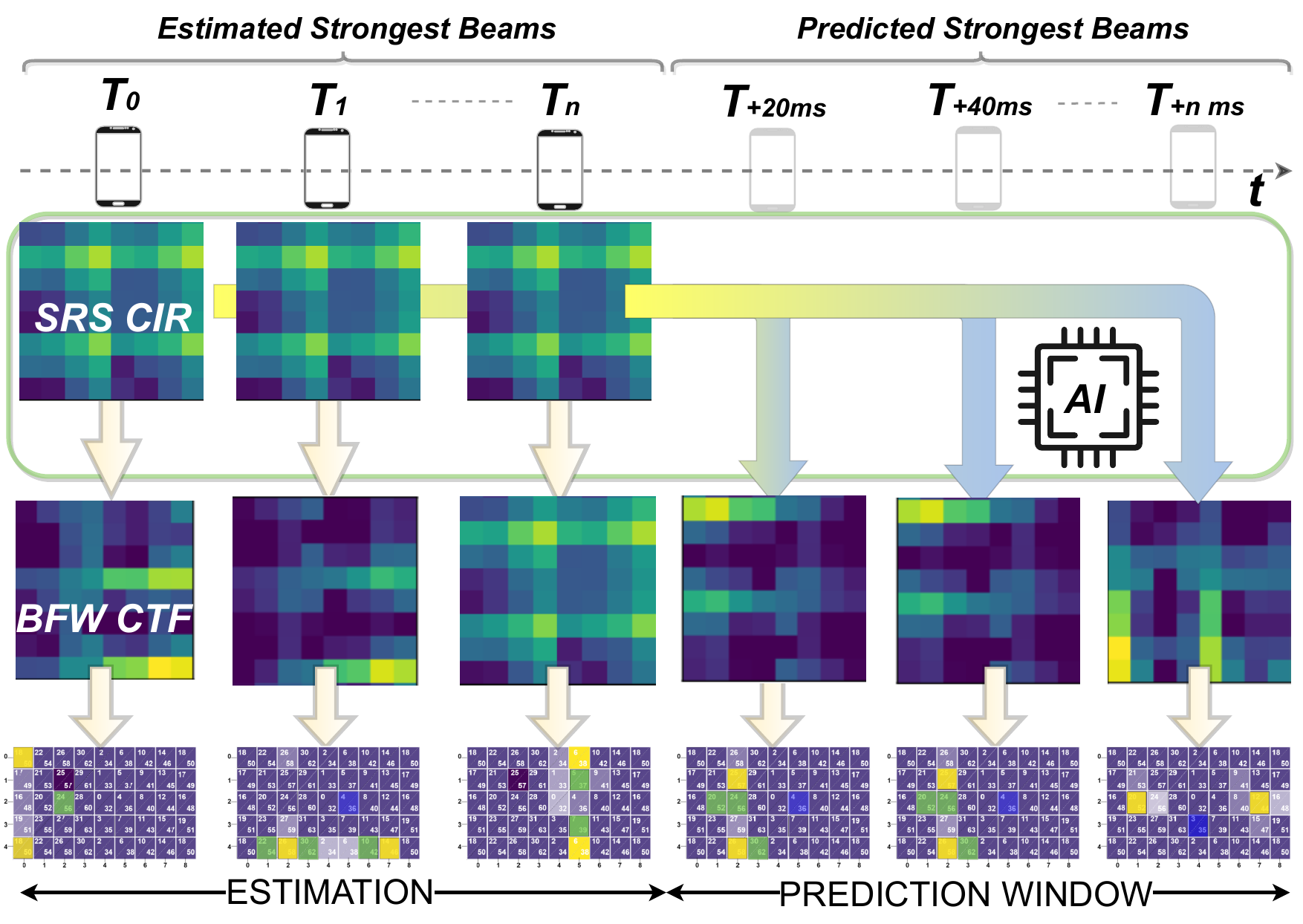}}
    \caption{Subsets of top-n beams are selected for future transmission from the predicted beam-domain set.}
    \label{fig:EstimationPredictionConcept}
\end{figure}
The energy of the best beams predicted by the suggested attention-based model, with its concept shown in Fig~\ref{fig:EstimationPredictionConcept}, is associated with the location of the UE. Determining the significance of the subset relative to the whole is valuable for beam prediction, which is highly related to the mobility of UE and scatterers. \newline
\indent In commercial deployments, threshold levels can be used to define the minimum beam subset size corresponding to the optimal energy levels, allowing the subset size to adapt dynamically rather than being restricted to the predefined discrete sizes considered in this study. \newline
\indent Statistical measures reduce extensive data sets to a single value, offering only one perspective on model errors by emphasizing specific aspects of model performance. To evaluate the performance of the proposed attention-assisted model, which uses time series data as input, we selected percentage-based error metrics such as the mean absolute percentage error (MAPE) (\ref{eqn:mape}) and the weighted mean absolute percentage error (WMAPE) (\ref{eqn:wmape}). MAPE quantifies the average magnitude of the error, while WMAPE, a variant of MAPE, adjusts the error calculations by incorporating real values or weights.
\begin{equation}
\begin{aligned}
   \label{eqn:mape} 
     \textrm{MAPE} = \frac{1}{n}\sum_{i=1}^n \frac{\mid y_i - \hat{y} \mid}{|y_i|} \times 100 \%
\end{aligned}
\end{equation}
\begin{equation}
\begin{aligned}
   \label{eqn:wmape} 
     \textrm{WMAPE} = \frac{\sum_{i=1}^{n} |y_i - \hat{y}_i|}{\sum_{i=1}^{n} |y_i|} \times 100 \%
\end{aligned}
\end{equation}

\begin{figure*}[htbp]
\begin{minipage}[t]{0.5\textwidth}
\hspace{-5pt}
\centering
\includegraphics[width=0.87\linewidth]{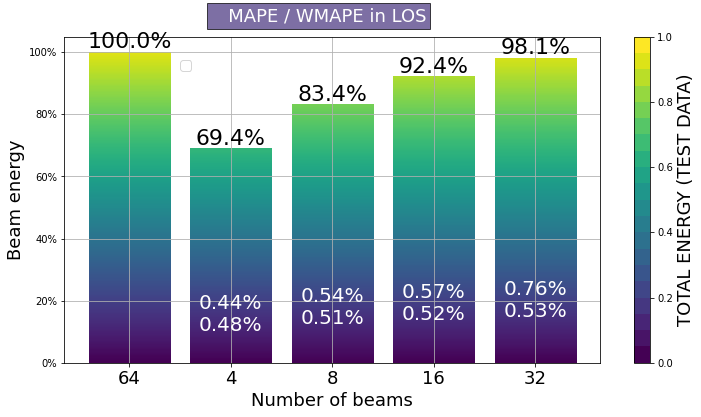}
\end{minipage}%
\hfill\allowbreak 
\begin{minipage}[t]{0.5\textwidth}
\hspace{-5pt}
\centering
\includegraphics[width=0.87\linewidth]{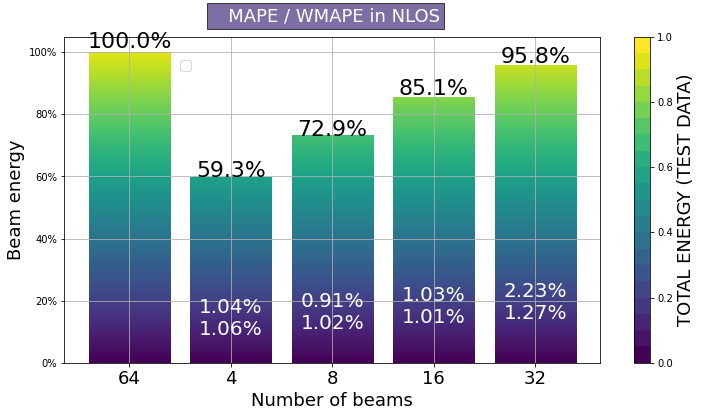}
\end{minipage}%
\caption{LoS/NLoS Comparison: Beam energy contributions of the estimated subsets relative to the total test data.}
\label{fig:EstimatedBeamPower}
\end{figure*}

\subsection{Evaluation of Estimation Results}
While the primary focus of this study is to enhance BM by predicting sets of best beams, we also underscore a significant capability of the proposed model: the accuracy of DL transfer functions estimation via BFWs generated by the MMSE algorithm.
As described earlier, BFWs are applied to the base station's MIMO antenna elements to shape and steer the signal by adjusting its amplitude and phase in the DL. Fig.~\ref{fig:EstimatedBeamPower} illustrates the high accuracy of the attention-based model in the LoS (right) and NLoS (left) scenarios. This result establishes a solid foundation for precise beam prediction in the DL. Furthermore, it highlights an important capability to generate BFWs, traditionally computed within the base station's baseband hardware, a process known for its intensive computational and hardware resource demands. These requirements are particularly challenging in multi-user MIMO scenarios, where the computational complexity grows linearly with the number of antennas at both the base station and user equipment.
\begin{figure*}[b!]
\begin{minipage}[t]{0.5\textwidth}
\hspace{-5pt}
\centering 
  \includegraphics[width=0.87\linewidth]{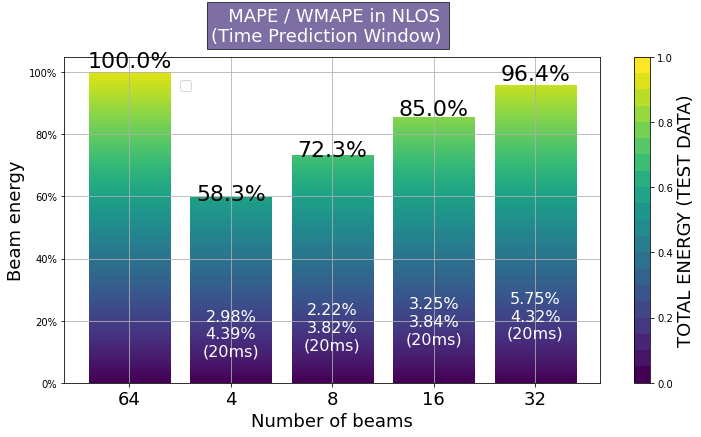}
\captionsetup{labelformat=empty}\addtocounter{figure}{-1}\caption{(a) 20ms, distance covered: 1.04$\lambda$}
\end{minipage}%
\hfill\allowbreak 
\begin{minipage}[t]{0.5\textwidth}
\hspace{-5pt}
\centering 
\includegraphics[width=0.87\linewidth]{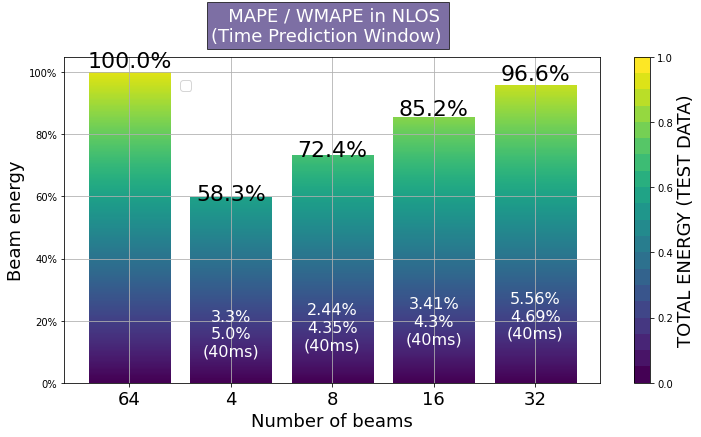}
\captionsetup{labelformat=empty}\addtocounter{figure}{-1}\caption{(b) 40ms, distance covered: 2.08$\lambda$}
\end{minipage}%
\hfill\allowbreak
\caption{NLoS comparison: Beam energy contributions of the predicted subsets within short prediction windows}
\label{fig:shortterm}
\end{figure*}
The NLoS results underscore the importance of focusing on a beam set, given the significant reduction in total energy caused by the dispersive characteristics of NLoS propagation. Moreover, it provides valuable information to establish suitable energy threshold levels to guide beam prediction strategies effectively.
\subsection{Evaluation of Prediction Results}
The attention-based model enables long-term prediction of best beam candidate subsets for UEs following predictable movement paths, leveraging historical SRS data transmitted by the UE in the UL. These predictions extend over timelines spanning multiple seconds, equivalent to several hundred wavelengths. This allows for simplifying the BM procedure for future time instances by allocating BM CSI-RS resources and configuring UE measurements solely for the beam subset identified by the prediction algorithm. 
As described in Section \ref{datacollection}, the NLoS scenario involves numerous scatterers with varying locations, sizes, and shapes, creating a high-dimensional feature space. These scatterers collectively influence the optimal beam direction, making it challenging to accurately model the complex scattering characteristics. However, the proposed model exhibits strong prediction accuracy, particularly for larger beam subsets, such as $n$ = 16, 32, ranked according to their expected power (e.g., strongest to weakest), the power in the weakest beams is expected to be very low compared to the cases of $n$ = 4, 8, as Fig. ~\ref{fig:shortterm} and Fig.~\ref{fig:longterm} illustrate. The latter observation is particularly emphasized in Fig.~\ref{fig:longterm} (d), where the attention model exhibits reduced prediction accuracy for larger beam subsets, such as 16 and 32 compared to Fig.~\ref{fig:longterm} (a) for instance. This outcome is expected, given the long-term perspective that extends well beyond the coherence time window, corresponding to hundreds of multiples of $\lambda$-wavelengths. However, since these findings enable beam prediction well in advance, they have the potential to fundamentally transform the way 5G networks manage BM resources.
\begin{figure*}[htbp]
\begin{minipage}[t]{0.5\textwidth}
\hspace{-5pt}
\centering
 \includegraphics[width=0.87\linewidth]{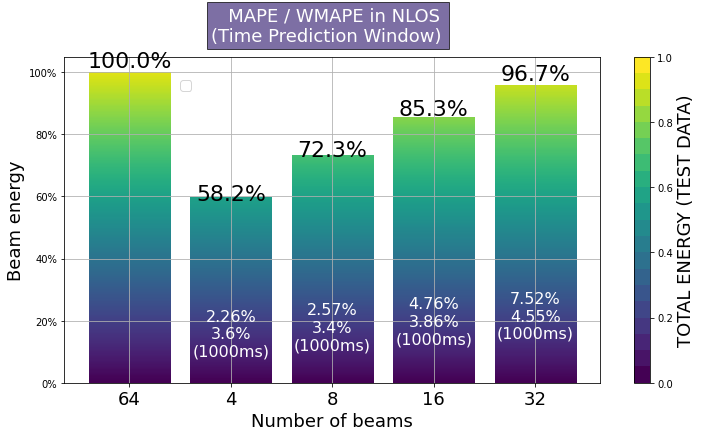}
\captionsetup{labelformat=empty}\addtocounter{figure}{-1}\caption{(a) 1000ms, distance covered: 52$\lambda$}
\end{minipage}%
\hfill\allowbreak
\begin{minipage}[t]{0.5\textwidth}
\hspace{-5pt}
\centering
 \includegraphics[width=0.87\linewidth]{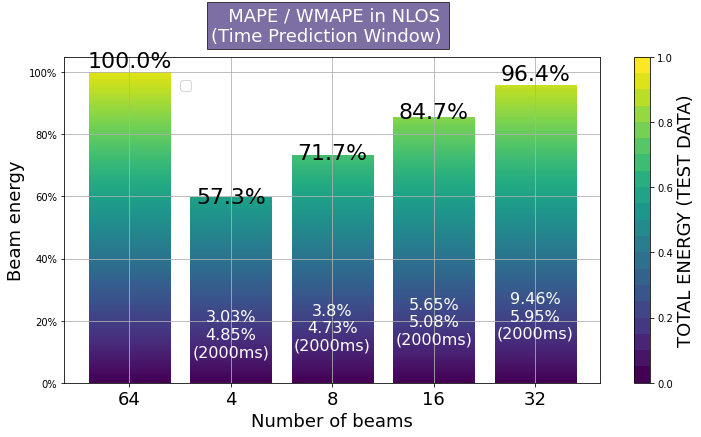}
\captionsetup{labelformat=empty}\addtocounter{figure}{-1}\caption{(b) 2000ms, distance covered: 104$\lambda$}
\end{minipage}%
\hfill\allowbreak
\begin{minipage}[t]{0.5\textwidth}
\hspace{-5pt}
\centering
 \includegraphics[width=0.87\linewidth]{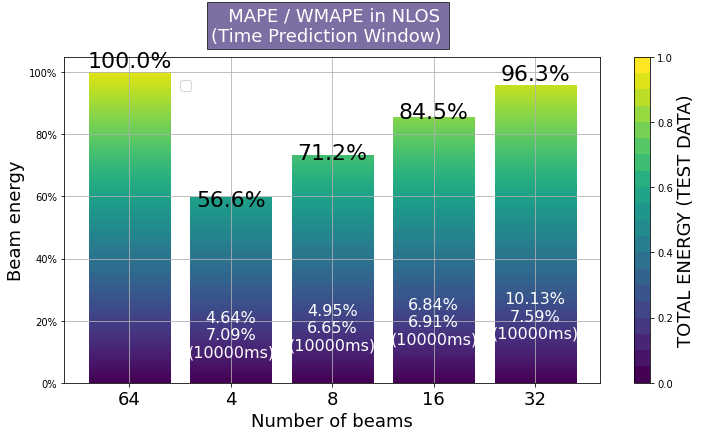}
\captionsetup{labelformat=empty}\addtocounter{figure}{-1}\caption{(c) 10000ms, distance covered: 521$\lambda$}
\end{minipage}%
\hfill\allowbreak
\begin{minipage}[t]{0.5\textwidth}
\hspace{-5pt}
\centering
 \includegraphics[width=0.87\linewidth]{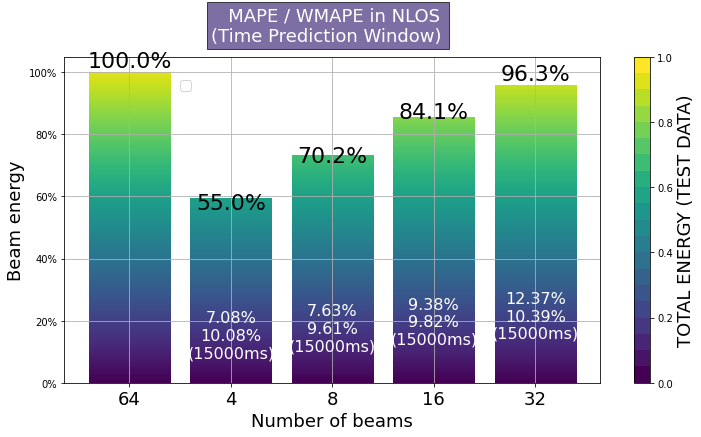}
\captionsetup{labelformat=empty}\addtocounter{figure}{-1}\caption{(d) 15000ms, distance covered: 781$\lambda$}
\end{minipage}%
\caption{NLoS comparison: Beam energy contributions of the predicted subsets within long prediction windows}
\label{fig:longterm}
\end{figure*}

\section{Conclusions}
\label{Conclusion}
We explored the application of transformers in massive MIMO beam prediction and demonstrated their competitive performance using datasets derived from commercial 5G systems. The proposed approach incorporates spatial and environmental factors, such as multipath scattering and predictable movement patterns, enabling the model to maintain high prediction accuracy even when temporal correlations diminish. This represents a significant advancement in 5G beam management, showcasing accurate beam prediction beyond coherence time, thereby overcoming the limitations of traditional methods constrained by coherence-time boundaries. The number of CSI-RS resources allocated for each BM measurement and the CSI-RS reporting rate can be significantly reduced compared to legacy BM operations. This reduction may enhance the DL spectral efficiency by preserving resources for data transmission and/or mitigating interference in the DL. In addition, it helps to meet the processing demands of the UE and improves the energy efficiency of the UE. \newline We acknowledge the repetitive nature of the selected UE trajectory and the UE antenna orientation toward the BS, as the UE was firmly mounted on the roof of the test vehicle. Although these factors reduced radio channel variations, the feasibility of attention-aided models for long-term beam prediction is still demonstrated. Another important consideration is the computational complexity of transformer models, which may pose a limitation. Thorough evaluation of their real-time performance is essential, and using fast GPUs or specialized hardware can help minimize delays, ensuring efficient operation in commercial deployments.



\end{document}